\documentclass[aps,pra,twocolumn,groupedaddress,showpacs]{revtex4}

\usepackage{latexsym,amsmath,amssymb}
\usepackage{graphicx}

\newcommand{\ket}[1]{
   \ensuremath{| #1 \rangle}}
\newcommand{\bra}[1]{
   \ensuremath{\langle #1 |}}
\newcommand{\braket}[2]{
   \ensuremath{\langle #1 | #2 \rangle}}

\begin{document}

%\preprint{}

\title{Dispersion cancellation and non-classical noise reduction for large photon-number states}

% Please use the appropriate macro for each type of information

\author{M. J. Fitch}
\email[]{michael.fitch@jhuapl.edu}
%\thanks{}
\author{J. D. Franson}
\email[]{james.franson@jhuapl.edu}

\affiliation{Johns Hopkins University, Applied Physics Laboratory,
Laurel MD 20723-6099}

\date{\today}

\begin{abstract}

Nonlocal dispersion cancellation is generalized to
frequency-entangled states with large photon number $N$. We show
that the same entangled states can simultaneously exhibit a factor
of $1/\sqrt{N}$ reduction in noise below the classical shot noise
limit in precise timing applications, as was previously suggested
by Giovannetti, Lloyd and Maccone (Nature \textbf{412} (2001)
417). The quantum-mechanical noise reduction can be destroyed by a
relatively small amount of uncompensated dispersion and entangled
states of this kind have larger timing uncertainties than the
corresponding classical states in that case. Similar results were
obtained for correlated states, anti-correlated states, and
frequency-entangled coherent states, which shows that these
effects are a fundamental result of entanglement.

\end{abstract}

\pacs{42.50.Dv, 03.67.-a, 03.65.-w}
%\keywords{}

\maketitle

%\section{}
%\subsection{}
%\subsubsection{}

%-----------------------------------------
\section{Introduction \label{sec:intro}}

     Two classical pulses of light propagating through two distant,
dispersive media will experience dispersion that depends only on
the local properties of the two media.  It was previously shown
\cite{Franson:1992}, however, that two entangled photons
propagating through two dispersive media can experience a nonlocal
cancellation of dispersion in the sense that the two photons will
arrive at two equally-distant detectors at the same time despite
the dispersion.  Other forms of dispersion cancellation have also
been discussed
\cite{SteinbergL:1992,SteinbergA:1992,Giovannetti:2001b,Larchuk:1995,%
Agarwal:1994,Jeffers:1993}.

     In this paper, we generalize dispersion cancellation to entangled
states containing a large number $N$ of photons in each pulse. We
also show that the same entangled states can simultaneously
exhibit a factor of $1/\sqrt{N}$ reduction in noise below the
classical shot noise limit in precise timing applications, such as
the synchronization of distant clocks
\cite{Jozsa:2000,Giovannetti:2001}.  We describe several different
examples of entangled states with large photon number that can
give this kind of behavior, including correlated states,
anti-correlated states, and entangled coherent states.

     An unexpected result of our analysis is that the $1/\sqrt{N}$ noise
reduction in timing measurements that was previously suggested by
Giovannetti, Lloyd, and Maccone \cite{Giovannetti:2001} can only
occur if there is very little dispersion to begin with or if
dispersion cancellation is used to reduce the effective dispersion
of the media.  Surprisingly little dispersion is required to
destroy the effect described in Ref.\ \cite{Giovannetti:2001}, and
the timing uncertainty from entangled states of this kind can
exceed the corresponding classical limit. As a result, the
potential effects of dispersion must be included when considering
non-classical noise reduction in precise timing applications, such
as clock synchronization. The use of quantum resources for clock
synchronization has also been discussed in Refs.\
\cite{Chuang:2000,Burt:2001,Jozsa:2001,Yurtsever:2000,Preskill:2000,Giovannetti:2002}.

Our main goal is to further consider the kinds of effects that can
be produced by entanglement in large photon-number states without
regard to whether or not the states of interest can be readily
produced using current experimental methods, as was the case in
Ref.\ \cite{Giovannetti:2001}. In some cases, however, we note
that the corresponding states can be experimentally produced for
small values of $N$.

%--------------------------------------------------------------------------
\section{Frequency Anti-Correlated 2N-photon state \label{sec:anti-corr-2N}}

\begin{figure}
\includegraphics[bbllx=1cm,bblly=12.5cm,bburx=16.5cm,bbury=20cm,width=3.2in]{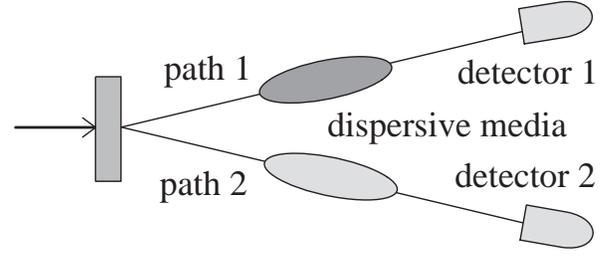}%
\caption{The source generates a state in which $N$ photons of
frequency $\omega_{o}+\epsilon$ travel in path 1 towards detector
1, and $N$ photons $(\omega_{o}-\epsilon)$ travel in path 2
towards detector 2. The dispersive medium in path 1 is described by
$\alpha_1$ and $\beta_1$, and similarly for path 2. \label{fig:2det}}
\end{figure}

The situation of interest is illustrated in Figure
(\ref{fig:2det}). Two non-classical beams of light propagate along
\mbox{paths} 1 and 2 through two dispersive media to two distant
detectors. The two beams are assumed to have a sufficiently small
bandwidth about a central frequency $\omega_{o}$ that the
dispersive properties of the two media can be characterized by
$k_{1}(\omega_{o}+\epsilon) = k_{o} + \alpha_{1} \epsilon +
\beta_{1} \epsilon^2$ and $k_{2}(\omega_{o}+\epsilon) = k_{o} +
\alpha_{2} \epsilon + \beta_{2} \epsilon^2$. Here $k_{1}(\omega)$
and $k_{2}(\omega)$
are the wave vectors %(wave numbers)
in the two media and $\alpha_{1}$, $\alpha_{2}$, $\beta_{1}$ and
$\beta_{2}$ are constants that represent the first few terms in a
Taylor series expansion. We will consider a single transverse
optical mode in each path, which could be approximated by a
single-mode optical fiber or by plane waves in free space. Our
goal is to consider the possibility of entangled states that can
eliminate the effects of dispersion while simultaneously reducing
the uncertainty in the difference of arrival times of the two
pulses below the classical shot noise limit.

In this section, we begin by considering entangled \mbox{states}
that contain $N$ photons in each path whose frequencies are
anti-correlated. Let $\ket{N(\omega)}_1$  and $\ket{N(\omega)}_2$
denote states with $N$ photons of frequency $\omega$ in path 1 or
path 2, respectively (Fock states). We then consider \cite{note1}
the state $\ket{\Psi}$ given by
\begin{equation}\label{eq:defPsi}
  \ket{\Psi}=\int_{-\infty}^{\infty} \! d\epsilon \, \phi(\omega_o + \epsilon) \,
  \ket{N(\omega_o + \epsilon)}_{1} \, \ket{N(\omega_o -
  \epsilon)}_{2}
\end{equation}
where $\phi(\omega)$ is a spectral function centered around
$\omega=\omega_{o}$. For $N=1$, $\ket{\Psi}$ can be produced by
spontaneous parametric down-conversion of a pump beam with
frequency $2\omega_o$, while Eq.\ (\ref{eq:defPsi}) is a
generalization to $N$ signal photons and $N$ idler photons. While
it is not currently known how to make such a state efficiently for
large values of $N$, the production of similar states with $N=2$
has been analyzed \cite{Weinfurter:2001} and demonstrated
\cite{Pan:2001}. (See also
\cite{Ou:1999,Lee:2001,Kok:2001,Fiurasek:2001,LLinares:2001}.)

By way of comparison, Giovannetti, Lloyd and Maccone
\cite{Giovannetti:2001} previously considered a state given by
\begin{equation}\label{eq:glm-state}
  \ket{\Psi}_{\mathrm{GLM}} = \int_{-\infty}^{\infty} \! d\epsilon
  \, \phi(\omega_o + \epsilon) \ket{N(\omega_o + \epsilon)}.
\end{equation}
They showed that the mean arrival time of the $N$ photons at a
single detector had an uncertainty that was below the classical
shot noise limit.  Equation (\ref{eq:glm-state}) differs from our
Eq.\ (\ref{eq:defPsi}) in that it involves a single photon mode
whereas Eq.\ (\ref{eq:defPsi}) includes two modes whose
frequencies are anti-correlated. As a result
$\ket{\Psi}_{\mathrm{GLM}}$ cannot give dispersion cancellation,
and the effects of dispersion were not included in the analysis of
Ref.\ \cite{Giovannetti:2001}.

The state $\ket{\Psi}$ can be written using photon creation
operators $\hat{a}_{1}^{\dagger}(\omega)$ and
$\hat{a}_{2}^{\dagger}(\omega)$ as:
\begin{equation}\label{eq:Psi-a-vac}
  \ket{\Psi} = \frac{1}{N!} \int \! d\epsilon \, \phi(\omega_o + \epsilon) \!
\left( \hat{a}^{\dagger}_1 (\omega_o + \epsilon) \right)^{\! N} \!
\left(\hat{a}^{\dagger}_2 (\omega_o - \epsilon) \right)^{\! N} \! \ket{0}
\end{equation}
where \ket{0} is the vacuum state.
Let the probability of detecting $N$ photons at times
$\{t_1,\cdots,t_N\}$ in detector 1 and $N$ photons at times
$\{t_{1}^{\prime},\cdots,t_{N}^{\prime}\}$ in detector 2  be
denoted $P(t_1,\cdots,t_N \, ; \,
t_{1}^{\prime},\cdots,t_{N}^{\prime} )$. The probability $P$ is
proportional to $\braket{A}{A}$ where the constant of
proportionality depends on the detection efficiency and $\ket{A}$
is defined by:
\begin{equation}\label{eq:def-A}
\begin{split}
  \ket{A(t_1,\cdots,t_N \, &; \, t_{1}^{\prime},\cdots,t_{N}^{\prime} )} =
  \hat{E}^{(+)}_1 (x,t_1) \cdots \hat{E}^{(+)}_1 (x,t_N) \\
  &\times \hat{E}^{(+)}_2 (x^{\prime},t_{1}^{\prime}) \cdots \hat{E}^{(+)}_2 (x^{\prime},t_{N}^{\prime})
  \ket{\Psi}. \\
\end{split}
\end{equation}
$\hat{E}^{(+)}_1 (x,t)$ is the positive frequency component of the
electric field operator and the distance from the source to the
detector in path 1 is assumed to be $x$, and $x^{\prime}$ for path
2.

The operators $\hat{E}^{(+)}_{1,2} (x,t_{j})$ can be expanded as
\begin{equation}\label{eq:E-FTa}
    \hat{E}^{(+)}_{1,2}(x,t_{j}) = \int^{\infty}_{0} \! d\omega \,
    \hat{a}_{1,2}(\omega) \, e^{i\left( k_{1,2}(\omega)x-\omega
    t_{j}\right) }
\end{equation}
where we have neglected a slowly varying function of $\omega$ and
we have suppressed dimensional constants. Using the commutator
$[\hat{a}_{\ell} (\omega_{i}) \, , \,
\hat{a}_{k}^{\dagger}(\omega_{j})] = \delta_{\ell k} \,
\delta(\omega_{i}-\omega_{j})$ \cite{note1} and combining Eqs.\
(\ref{eq:def-A}) and (\ref{eq:E-FTa}) gives:
\begin{equation}\label{eq:A-1}
\begin{split}
  \ket{A} &= \frac{1}{N!} \int d\epsilon \, \phi(\omega_o + \epsilon) \\
    &\!\!\times \exp[i( k_{1}(\omega)x-\omega t_{1})] \cdots
            \exp[i( k_{1}(\omega)x-\omega t_{N})] \\
    &\!\!\times \exp[i( k_{2}(\omega^{\prime})x^{\prime}\!-\omega^{\prime} t_{1}^{\prime})] \cdots
            \exp[i( k_{2}(\omega^{\prime})x^{\prime}\!-\omega^{\prime} t_{N}^{\prime})] \ket{0} \\
\end{split}
\end{equation}
where $\omega \equiv \omega_o + \epsilon$ and $\omega^{\prime}
\equiv \omega_o - \epsilon$.

%%%%%%%%%%%%%%%%%%%%%%%%%%%%%%%%%%%%%%%%%%%%%%%%%%%%%%%%%%%%%%
%%%%%%%%%%%%%%%%%%%%%%%%%%%%%%%%%%%%%%%%%%%%%%%%%%%%%%%%%%%%%%

Including the dispersive properties of the two media, the
amplitude $A$ of state \ket{A} can be written as
\begin{equation}\label{eq:A-2}
\begin{split}
  A &= \frac{1}{N!} \int d\epsilon \, \phi(\omega_o + \epsilon)
  \exp[i \epsilon N (\alpha_1 x - \alpha_2 x^{\prime})] \\
  &\times \exp[i \epsilon^2 N (\beta_1 x + \beta_2 x^{\prime})]
  \exp[-i \epsilon N(\overline{t} - \overline{t}^{\prime} )]. \\
\end{split}
\end{equation}
An overall phase factor has been dropped and we have defined the
mean detection times $\overline{t} \equiv \frac{1}{N} \sum_{j=1}^N
t_j$ and similarly $\overline{t}^{\prime} \equiv \frac{1}{N}
\sum_{j=1}^N t_j^{\prime}$.

It can be seen from Eq.\ (\ref{eq:A-2}) that the effects of
dispersion will cancel non-locally between the two media if
$\beta_{1} x= -\beta_{2} x^{\prime}$, as was shown previously
\cite{Franson:1992} for the case of $N=1$. The non-classical noise
reduction inherent in Eq.\ (\ref{eq:A-2}) can be best understood
by comparing these results with those from a single-photon wave
packet with the same spectral function $\phi(\omega)$:
\begin{equation}\label{eq:def-single-Psi}
  \ket{\Psi_{1}} = \int_{-\infty}^{\infty} \! d\epsilon \,
  \phi(\omega_{o} + \epsilon) \hat{a}^{\dagger} \ket{0}
\end{equation}
If we define $g(t)$ as the Fourier transform of $\phi(\omega)$
\begin{equation}\label{eq:def-gt}
  g(t) = \int_{-\infty}^{\infty} \! \phi(\omega) e^{-i \omega t} \,
  d\omega
\end{equation}
then $|g(t)|^2$ gives the intensity of the wave packet as a
function of time at a fixed position ($x=0$). A similar result
holds for classical pulses, and the width $\sigma_{g}$ of
$|g(t)|^2$ is equal to the uncertainty in the timing information
that can be obtained from a single photon wave packet with
spectral function $\phi(\omega)$. If $N$ independent single-photon
wave packets are transmitted, the timing uncertainty can be
reduced to $\sigma_{g}/\sqrt{N}$, which corresponds to the
classical shot noise limit. Comparison with Eq.\ (\ref{eq:A-2})
for the case in which $x=x^{\prime}$, $\alpha_{1}=\alpha_{2}$, and
$\beta_{1} = -\beta_{2}$ shows that, for the entangled state
\ket{\Psi}
\begin{equation}\label{eq:gNtau}
  |A(\tau)|^2 = \frac{1}{N!^2} |g(N\tau)|^{2}.
\end{equation}
Here $\tau$ is defined as the difference in mean arrival times,
$\tau = \overline{t} - \overline{t}^{\prime}$. It can be seen from
Eq.\ (\ref{eq:gNtau}) that $|A|^2$ has a width that is a factor of
$1/N$ less than that of the corresponding single-photon wave
packet, which results in a factor of $1/\sqrt{N}$ reduction in the
noise as compared to $N$ independent single-photon wave packets.
Thus our results with dispersion cancellation are similar to those
obtained from Eq.\ (\ref{eq:glm-state}) by Giovannetti, Lloyd, and
Maccone \cite{Giovannetti:2001}, who did not include the effects
of dispersion. The factor of $1/N$ improvement in timing
resolution is closely related to the $1/N$ improvement in spatial
resolution proposed by Boto \emph{et al.}\ for use in quantum
lithography \cite{Boto:2000}.

     The above results show that nonlocal cancellation of
dispersion and a non-classical reduction of noise can occur
simultaneously for entangled states with large photon numbers. The
condition $\beta_{1} x= - \beta_{2} x^{\prime}$ can be achieved in
optical fibers \cite{Brendel:1998}, for example, but would be
difficult to achieve under more general conditions, such as for
the case of light beams propagating through the atmosphere. As a
result, it is important to consider the effects of uncompensated
dispersion on the non-classical noise reduction of states of this
kind.  All of the necessary integrals can be evaluated
analytically for the case in which the spectral function
$\phi(\omega_o + \epsilon)$ is assumed to be Gaussian with width
$\sigma_{\phi}$
\begin{equation}\label{eq:def-phi}
  \phi(\omega_o +\epsilon)=\exp(-\epsilon^2 /2 \sigma_{\phi}^2).
\end{equation}
For simplicity, we have omitted a normalization constant in Eq.\
(\ref{eq:def-phi}). A Gaussian may, for example, represent a
narrow bandwidth filter of the kind that is widely used in
down-conversion experiments. For convenience, we will define
$a_{\phi}^2 \equiv (2 \sigma_{\phi}^2)^{-1}$, and obtain
\begin{equation}\label{eq:A-3}
  A=\frac{1}{N!} \int_{-\infty}^{\infty} \!\! d\epsilon \,
  \exp\{-[ \epsilon^2 ( a_{\phi}^2 - i N B) + i \epsilon N
  \zeta]\}
\end{equation}
where we have defined $B \equiv (\beta_1 x + \beta_2 x^{\prime})$
and $\zeta \equiv \tau-(\alpha_1 x - \alpha_2 x^{\prime})$.
(Recall $\tau = \overline{t} - \overline{t}^{\prime}$.)

The integral may be evaluated by completing the square in the
exponent, and aside from an overall phase factor
\begin{equation}\label{eq:A-4}
  A=\frac{1}{N!} \frac{\sqrt{\pi}}{\sqrt{\smash[b]{a_{\phi}^2 - i
  N B}}} \exp\left(\frac{-N^2 \zeta^2}{4 (a_{\phi}^2 - i
  N B)} \right)
\end{equation}
%%%%%%%%%%%%%%%%%%%%%%% <A|A> = |A|^2 = ... %%%%%%%%%%%%%%%%%%%%
\begin{equation}\label{eq:A-5}
  \braket{A}{A}\! =\!|A|^2 \!=\!
  \frac{\pi}{N!^{2} \sqrt{\smash[b]{a_{\phi}^4 + N^2 B^2}}}
  \exp\!\left(\frac{-(\tau-\overline{\tau})^2 N^2 a_{\phi}^2}{2(a_{\phi}^4 + N^2
  B^2)} \right)
\end{equation}
where $\overline{\tau} = (\alpha_1 x - \alpha_2 x^{\prime})$.
Equation (\ref{eq:A-5}) is thus a Gaussian in $\tau$ with mean
$\overline{\tau}$ and width $\sigma_{\tau}$ given by
\begin{equation}\label{eq:A-6-sigma}
  \sigma_{\tau}^2 = \frac{ a_{\phi}^4 + N^2 B^2}{N^2
  a_{\phi}^2}
\end{equation}
\begin{equation}\label{eq:A-7-sigma}
  \sigma_{\tau}^2 = \frac{ 1 + 4 \sigma_{\phi}^4 N^2 (\beta_1 x +
  \beta_2 x^{\prime})^2}{2 \sigma_{\phi}^2 N^2}
\end{equation}
If the distances $x$ and $x^{\prime}$ are regarded as known, then
$|A|^2$ describes a Gaussian probability distribution for the
difference in mean arrival times. As we observed above, this
allows dispersion cancellation \cite{Franson:1992} for $\beta_1 x
= - \beta_2 x^{\prime}$, in which case $\sigma_{\tau} =
1/(\sqrt{2} \sigma_{\phi} N)$.

However, if the dispersion cannot be cancelled or neglected, the
scaling with large $N$ is independent of $N$:
\begin{equation}\label{eq:lim-A-7}
  \lim_{N\rightarrow\infty} \sigma_{\tau} = \sqrt{2} \sigma_{\phi}
  \left| \beta_1 x + \beta_2 x^{\prime} \right|
\end{equation}
One might expect to approach the classical limit for large photon
numbers and indeed in this case the quantum mechanical enhancement
disappears. In fact, the timing uncertainty associated with the
entangled state $\ket{\Psi}$ under these conditions is worse than
the classical shot noise limit, since Eq.\ (\ref{eq:lim-A-7}) does
not include a factor of $1/\sqrt{N}$, as is the case classically.
The transition point between these two limits occurs when
\begin{equation}\label{eq:trans-A-7}
  1 = 4 \sigma_{\phi}^4 N_{\mathrm{transition}}^2 (\beta_1 x + \beta_2
  x^{\prime})^2
\end{equation}
At the transition point, $\sigma_{\tau}$ is $\sqrt{2}$ times the
limiting value of Eq.\ (\ref{eq:lim-A-7}), and further increases
of $N$ above $N_{\mathrm{transition}}$ have diminishing effects.
For typical optical materials such as fused silica, the group
delay dispersion (at 800~nm) is $2\beta \sim 500\ %
\mathrm{fs}^2/\mathrm{cm}$ \cite{Walmsley:2001}. For a 5~nm
bandwidth (one sigma) centered at 800~nm, $\sigma_{\phi} \sim 3.7 \times %
10^{11}$\ rad/sec. A 1~cm thickness of fused silica in both paths
would produce enough dispersion such that $N_{\mathrm{transition}}
\sim 7.3 \times 10^{3}$.  Choosing a more modest
$N_{\mathrm{transition}}=100$, the dispersion of 146~cm of fused
silica in both paths is required. Similar remarks apply to the
state vector of Eq.\ (\ref{eq:glm-state}), which does not allow
dispersion cancellation.

Non-classical noise reduction of this kind has been proposed
\cite{Giovannetti:2001} for the synchronization of clocks on
orbiting satellites. But in applications of that kind, the
atmospheric dispersion cannot in general be cancelled or neglected
with the exception of satellite-to-satellite links that do not
pass through the atmosphere. We can estimate the dispersion of air
using empirical relations for
the index of refraction. Since $k_{\mathrm{air}}(\omega) = \frac{\omega}{c} %
n_{\mathrm{air}}(\omega)$ we have
\begin{equation}\label{eq:def-beta}
  \beta = \frac{1}{2}\, \frac{d^2 k(\omega)}{d\omega^2} =
  \frac{1}{2c} \, \frac{d^2}{d\omega^2}\left( \omega \, n(\omega)\right)
\end{equation}
The widely-used $n_{\mathrm{air}}$ formula of Edl\'{e}n
\cite{Edlen:1966} yields $\beta_{\mathrm{air}}= 0.106$\ fs$^2$/cm
at 800~nm for standard dry air at 15$^{\circ}$C. Including 20\%
relative humidity using the more accurate formula of Owens
\cite{Owens:1967} for $n_{\mathrm{air}}$ gives
$\beta_{\mathrm{air}} = 0.103$\ fs$^2$/cm at 800~nm. Thus
$\sim$24~m of air has dispersion equivalent to 1~cm of fused
silica, and long air path lengths would significantly limit the
non-classical noise reduction suggested in Ref.\
\cite{Giovannetti:2001}.

\begin{figure}
\includegraphics[width=3.2in]{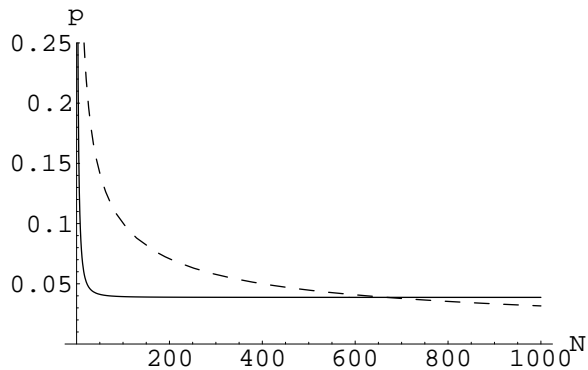}%
\caption{A comparison of the classical (dashed line) and quantum
(solid line) timing uncertainties. The widths
$\sigma_{\mathrm{Q}}$ and $\sigma_{\mathrm{C}}$ have been
multiplied by the spectral bandwidth $\sigma_{\phi}$ to give a
dimensionless parameter $p$. Note that for a fixed dispersion, (in
this case 4~m of fused silica or 10~km of air) the quantum
mechanical expression rapidly approaches its asymptote, while the
corresponding classical width decreases as $1/\sqrt{N}$ from
classical averaging.
 \label{fig:air10km}}
\end{figure}

For a pair of classical Gaussian pulses, it was previously shown
\cite{Franson:1992} that
\begin{equation}\label{eq:classical-1}
  |g(\tau)|^2 = \exp(-(\tau-\tau_o)^{2} /2 \sigma_{T}^2)
\end{equation}
where $\tau_{o}=\alpha_2 x_2 - \alpha_1 x_1$, and the width, in
the notation used here, is
\begin{equation}\label{eq:classical-2}
  \sigma_{T}^2 = \frac{2 a_{\phi}^4 + (\beta_1^2 x_1^2 + \beta_2^2 x_2^2)}
  {a_{\phi}^2}.
\end{equation}
There is no possibility of dispersion cancellation in the
classical case since Eq.\ (\ref{eq:classical-2}) contains the sum
of squares of the $\beta$ coefficients. In the limit of large
dispersion, it can be seen that Eq.\ (\ref{eq:classical-2}) is
equivalent to the limiting case of Eq.\ (\ref{eq:lim-A-7}), except
that the overall timing uncertainty in Eq.\ (\ref{eq:classical-2})
can be reduced by the square root of the number of photons in each
path by classical averaging. This gives the classical timing
uncertainty $\sigma_{\mathrm{C}} \equiv \sigma_{T}/\sqrt{N}$. In
the quantum case, Eqs.\ (\ref{eq:A-7-sigma}) and
(\ref{eq:lim-A-7}) already represent the distribution of the
average detection time, giving the quantum timing uncertainty
$\sigma_{\mathrm{Q}} \equiv \sigma_{\tau}$. The quantum and
classical expressions are plotted as a function of $N$ in Fig.\
\ref{fig:air10km} for a dispersive path of 4~m of fused silica (or
10~km of air), and a bandwidth $\sigma_{\phi}=3.7 \times 10^{11}$\
Hz centered at 800~nm. We plot $\sigma_{\phi}\sigma_{\mathrm{Q}}$,
which is dimensionless, and similarly $\sigma_{\phi}
\sigma_{\mathrm{C}}$ for the classical expression.

\begin{figure}
\includegraphics[width=3.2in]{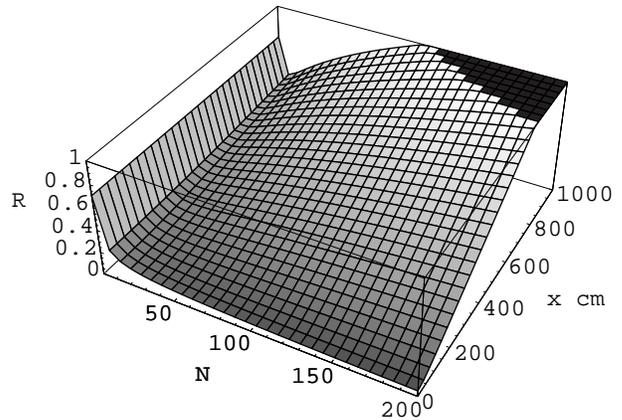}%
\caption{A surface plot of the ratio $R$ of the quantum mechanical
width to the classical width,
$\sigma_{\mathrm{Q}}/\sigma_{\mathrm{C}}$, as a function of $N$
and propagation distance $x$ (cm) through a dispersive medium with
$\beta=250$\ fs$^{2}$/cm (fused silica) and a bandwidth
$\sigma_{\phi}=3.7 \times 10^{11}$\ Hz. The quantum mechanical
timing uncertainty is larger than the corresponding classical case
in the black region in the upper right hand corner.
\label{fig:surfaceplot}}
\end{figure}

When the photon number or the dispersion is large, the quantum
mechanical timing uncertainty is larger than the corresponding
classical case, as shown in Fig.\ \ref{fig:surfaceplot}, where the
ratio $\sigma_{\mathrm{Q}} / \sigma_{\mathrm{C}}$ is plotted as a
function of photon number $N$ and distance $x$ (cm) in fused
silica. To highlight the region where the quantum timing
uncertainty is larger than the classical timing uncertainty, the
plot has been clipped at unity and the clipped region rendered
black.

%%%%%%%%%%%%%%%%%%%%%%%%%%%%%%%%%%%%%%%%%%%%%%%%%%%%%%%%%%%%%%
%%%%%%%%%%%%%%%%%%%%%%%%%%%%%%%%%%%%%%%%%%%%%%%%%%%%%%%%%%%%%%

%%%%%%%%%%%%%%%%%%%%%%%%%%%%%%%%%%%%%%%%%%%%%%%%%%%%%%%%%%%%%%
%%%%%%%%%%%%%%%%%%%%%%%%%%%%%%%%%%%%%%%%%%%%%%%%%%%%%%%%%%%%%%

%------------------------------------------------
\section{Thick detectors \label{sec:thick}}

\begin{figure*}
\includegraphics[bbllx=1.4cm,bblly=15.5cm,bburx=19.7cm,bbury=20cm,width=6.4in]{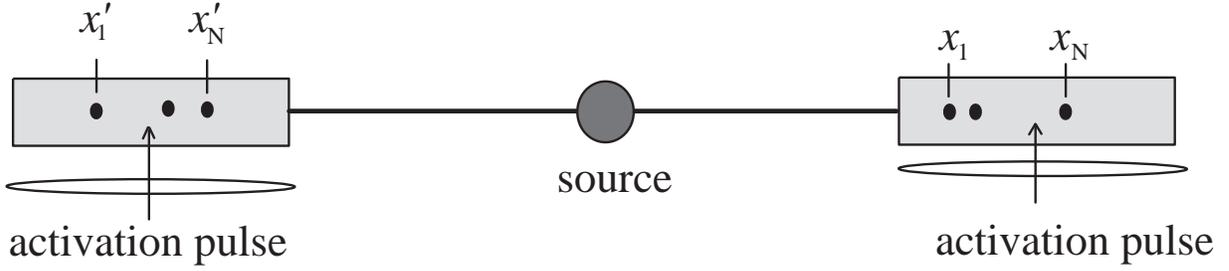}%
\caption{A thick, position-sensitive detector is assumed to be
gated on for a narrow time interval $\delta t$ by a laser pulse.
The positions of the detected photons, which are $\{x_1 \cdots
x_N\}$ for one side and $\{x_1^{\prime} \cdots x_N^{\prime} \}$
for the other, are recorded by the detectors. \label{fig:thick}}
\end{figure*}

In the previous section, we considered the distance from the
source to the detectors as fixed and found that the spread in
arrival times had a narrow distribution. What if the roles of $x$
and $t$ are interchanged at the detector?  Here we consider a
\emph{gedanken} experiment in which the detectors are thick in the
$x$-direction, and can be activated (gated on) for a narrow time
interval $\delta t$. During the time that it is gated on, the
photons travelling inside a detector have some probability of
being detected at positions that are registered. Such a detector
might resemble a photographic emulsion that is activated by an
ultrashort laser pulse, as depicted in Fig.\ \ref{fig:thick}. We
now calculate the distribution of the detected positions.

%%%%%%%%%%%%%%%%%%%%%%%%%%%%%%%%%%%%%%%%%%%%%%%%%%%%%%%%%%%%%%%%%%
%%%%%%%%%%%%%%%%%%%%%%%%%%%%%%%%%%%%%%%%%%%%%%%%%%%%%%%%%%%%%%%%%%

Considering the same entangled state $\ket{\Psi}$ as before in
Eqs.\ (\ref{eq:defPsi}) and (\ref{eq:Psi-a-vac}), we now consider
the detection time in each arm fixed and ask what are the
positions of the photons at the time of detection. Here we
construct $A=A(x_1 \cdots x_N ; x_1^{\prime} \cdots x_N^{\prime} ;
t , t^{\prime}, \delta t)$, where $|A|^2$ is proportional to the
probability of detecting $N$ photons at positions $x_1 \, , \cdots
, \, x_N$ at detector 1 in time interval $(t_1, t_1 + \delta t)$
and $N$ photons at positions $x_1^{\prime} \, ,  \cdots  , \,
x_N^{\prime}$ at detector 2 in time interval $(t_2, t_2 + \delta
t)$.
\begin{equation}\label{eq:thickA-1}
\begin{split}
  \ket{A} = \int_{t_{1}}^{t_{1} + \delta t} \!\! dt \int_{t_{2}}^{t_{2} + \delta t} \!\! dt^{\prime}
  & \hat{E}^{(+)}_1 (x_1 , t) \cdots \hat{E}^{(+)}_1 (x_N , t) \\
  \times & \hat{E}^{(+)}_2 (x_1^{\prime} , t^{\prime}) \cdots \hat{E}^{(+)}_2 (x_N^{\prime} , t^{\prime}) \ket{\Psi} \\
\end{split}
\end{equation}
Making use of the expansion of the field operators in Eq.\
(\ref{eq:E-FTa}) and the commutator as before gives the
probability amplitude A:
\begin{equation}\label{eq:thickA-2}
\begin{split}
  A &= \frac{1}{N!} \int \! dt \int \! dt^{\prime}
  \int d\epsilon \, \phi(\omega_o + \epsilon) \\
  &\times \exp[i\left( k(\omega)x_1 -\omega t\right)] \cdots
          \exp[i\left( k(\omega)x_N -\omega t\right)] \\
  &\times \exp[i\left( k(\omega^{\prime})x_1^{\prime}-\omega^{\prime} t^{\prime}\right)] \cdots
          \exp[i\left( k(\omega^{\prime})x_N^{\prime}-\omega^{\prime} t^{\prime}\right)] \\
\end{split}
\end{equation}
where again $\omega=\omega_o + \epsilon$ and
$\omega^{\prime}=\omega_o - \epsilon$. We define $\overline{x}
\equiv (1/N) \sum_{j=1}^N x_j$ and similarly
$\overline{x}^{\prime} \equiv (1/N) \sum_{j=1}^N x_j^{\prime}$.
Including the dispersive properties of the media gives:
\begin{equation}\label{eq:thickA-3}
\begin{split}
  A &= \frac{1}{N!} \int \!\! dt \int \!\! dt^{\prime}
  \int \!\! d\epsilon \, \phi(\omega_o + \epsilon) \\
  &\times \exp i \epsilon N (\alpha_1 \overline{x} - \alpha_2 \overline{x}^{\prime})
          \exp i \epsilon^2 N (\beta_1 \overline{x} + \beta_2 \overline{x}^{\prime})  \\
  &\times \exp (-i \omega_o N (t+t^{\prime})) \exp(-i \epsilon N (t-t^{\prime}))  \\
\end{split}
\end{equation}
where an overall phase factor has been omitted. This can be
rearranged to give
\begin{equation}\label{eq:thickA-4}
\begin{split}
  A &= \frac{1}{N!} \int_{t_{1}}^{t_{1} + \delta t} \!\!\! dt
                              \int_{t_{2}}^{t_{2} + \delta t} \!\!\! dt^{\prime}
           \exp[-i \omega_o N (t+t^{\prime})] \\
    &\times \int \! d\epsilon \, \phi(\omega_o + \epsilon)
     \exp [i \epsilon N ((\alpha_1 \overline{x}-t) - (\alpha_2
       \overline{x}^{\prime}-t^{\prime}))] \\
    &\times \exp [i \epsilon^2 N (\beta_1 \overline{x} + \beta_2 \overline{x}^{\prime})].  \\
\end{split}
\end{equation}
Note that the quantity $(\alpha_1 \overline{x}-t)$ corresponds to
the group delay along path 1, and a similar factor appears for
path 2.

In the limit $\delta t \ll 1/(N \omega_o)$, the time integrations
become trivial and the factor $\exp (-i \omega_o N
(t+t^{\prime}))$ reduces to a constant phase shift. We define
$\tau \equiv t -t^{\prime}$ and $\tau_o \equiv \alpha_1
\overline{x} - \alpha_2 \overline{x}^{\prime}$.  Then Eq.\
(\ref{eq:thickA-4}) is proportional to:
\begin{equation}\label{eq:thickA-5}
  A = \frac{1}{N!} \int \!\! d\epsilon \, \phi(\omega_o + \epsilon)
  e^{-i \epsilon N (\tau - \tau_o)}
  e^{i \epsilon^2 N (\beta_1 \overline{x} + \beta_2 \overline{x}^{\prime})}.
\end{equation}
Assuming a Gaussian for $\phi(\omega_o + \epsilon)$ as in Eq.\
(\ref{eq:def-phi}), we can complete the square in the exponent to
obtain
\begin{equation}\label{eq:thickA-6}
  A = \frac{1}{N!} \frac{\sqrt{\pi}}{\sqrt{\smash[b]{a_{\phi}^2 -i NB}}}
  \exp \left(- \frac{N^2 (\tau -\tau_o)^2}{4 (a_{\phi}^2 -iNB)} \right)
\end{equation}
\begin{equation}\label{eq:thick-absA-1}
  |A|^2 = \frac{1}{N!^2} \frac{\pi}{\sqrt{\smash[b]{a_{\phi}^4 + N^2 B^2}}}
  \exp \left( - \frac{ (\tau - \tau_o)^2 N^2 a_{\phi}^2}{2( a_{\phi}^4 + N^2 B^2)} \right)
\end{equation}
This has the same features as Eq.\ (\ref{eq:A-5}) and corresponds
to a Gaussian in the variable $\tau = t - t^{\prime}$ with mean
$\tau_o \equiv \alpha_1 \overline{x} - \alpha_2
\overline{x}^{\prime}$. The width is the same as in Eq.\
(\ref{eq:A-6-sigma}) and (\ref{eq:A-7-sigma}).

Equation (\ref{eq:thick-absA-1}) shows that a gated detector can
exhibit dispersion cancellation and nonlocal noise reduction
involving the average position of detection that is analogous to
the average detection times from a more conventional detector. If
$\alpha_1 = \alpha_2$, then the mean positions will be very nearly
equal if $t_1 =t_2$. If the clocks are known to be synchronized,
then any difference between the distances to the two detectors can
be found. Similarly, if the distances to the detectors are known,
then the synchronization of the clocks could be checked.

%------------------------------------------------------
\section{Correlated 2N-photon state \label{sec:corr-2N}}

The entangled state \ket{\Psi} of Eqs.\ (\ref{eq:defPsi}) and
(\ref{eq:Psi-a-vac}) corresponds to two beams of light with
anti-correlated frequencies, as is produced by parametric
down-conversion for the case of $N=1$. In this section, we
investigate the question of whether or not nonlocal cancellation
of dispersion and non-classical noise reduction can occur for an
entangled state with correlated frequencies instead, such as the
state \ket{\Psi^{\prime}} given by
\begin{equation}\label{eq:def-psiprime}
    \ket{\Psi^{\prime}}=\int d\epsilon \, \phi(\omega_o + \epsilon) \,
  \ket{N(\omega_o + \epsilon)}_{1} \, \ket{N(\omega_o +
  \epsilon)}_{2}
\end{equation}
Newly proposed techniques for source engineering
\cite{Branning:1999,Erdmann:2000,Banaszek:2001} may be able to
produce the state $\ket{\Psi^{\prime}}$ at least for small values
of $N$. States with spectral correlation and anti-correlation were
theoretically studied (for the case $N=1$) by Campos \emph{et
al.}\ \cite{Campos:1990}.

As above, the probability of detecting $N$ photons at times
$\{t_1,\cdots,t_N\}$ and positions $\{x_1,\cdots,x_N\}$ in
detector 1 and $N$ photons at times
$\{t_{1}^{\prime},\cdots,t_{N}^{\prime}\}$ and positions
$\{x_{1}^{\prime},\cdots,x_{N}^{\prime}\}$ in detector 2 is
proportional to an amplitude $|A^{\prime}|^2$ given by
\begin{equation}\label{eq:Aprime-1}
\begin{split}
  A^{\prime} &= \frac{1}{N!} \int d\epsilon \, \phi(\omega_o + \epsilon) \\
  &\times \exp[i\left( k(\omega)x_1-\omega t_{1}\right)] \cdots
          \exp[i\left( k(\omega)x_N-\omega t_{N}\right)] \\
  &\times \exp[i\left( k(\omega^{\prime})x_1^{\prime}-\omega^{\prime} t_{1}^{\prime}\right)] \cdots
          \exp[i\left( k(\omega^{\prime})x_N^{\prime}-\omega^{\prime} t_{N}^{\prime}\right)] \\
\end{split}
\end{equation}
where here, $\omega= \omega^{\prime} = \omega_{o} + \epsilon$.
Including the effects of a dispersive medium in paths 1 and 2 and
collecting the terms gives:
\begin{equation}\label{eq:Aprime-2}
\begin{split}
  A^{\prime} &= \frac{1}{N!} \! \int \!\! d\epsilon \, \phi(\omega_o + \epsilon)
    \exp[ i \epsilon N (\alpha_1 \overline{x}+ \alpha_2
    \overline{x}^{\prime})] \\
   &\times \exp[ i \epsilon^2 N (\beta_1 \overline{x} + \beta_2
    \overline{x}^{\prime})]
    \exp[- i \epsilon N (\overline{t} + \overline{t}^{\prime})] \\
\end{split}
\end{equation}
As before, we assume that $\phi(\omega_o + \epsilon)$ is a
Gaussian, and for convenience, we define $B \equiv (\beta_1
\overline{x} + \beta_2 \overline{x}^{\prime})$ and $\xi \equiv
(\overline{t}-\alpha_1 \overline{x} + \overline{t}^{\prime} -
\alpha_2 \overline{x}^{\prime})$. Then Eq.\ (\ref{eq:Aprime-2})
becomes:
\begin{equation}\label{eq:Aprime-3}
  A^{\prime} =\frac{1}{N!} \int_{-\infty}^{\infty}
  d\epsilon \exp \bigl(-[\epsilon^2 (a_{\phi}^{2} - i N B) + i \epsilon
  N \xi]\bigr).
\end{equation}
This integral can be evaluated to give
\begin{equation}\label{eq:Aprime-4}
  A^{\prime} =\frac{1}{N!}  \frac{\sqrt{\pi}}{\sqrt{\smash[b]{a_{\phi}^2 -i
  NB}}} \exp\left(-\frac{N^2 \xi^2}{4 (a_{\phi}^2 -i
  NB)} \right)
\end{equation}
\begin{equation}\label{eq:Aprime-abs-5}
  |A^{\prime}|^2 = \frac{1}{N!^2} \frac{\pi}{ \sqrt{\smash[b]{a_{\phi}^4 +N^2
  B^2}}} \exp\left( -\frac{(\tau^{\prime} - \overline{\tau}^{\prime})^2 N^2 a_{\phi}^2}{2(a_{\phi}^4 + N^2
  B^2)}\right)
\end{equation}
In going from Eq.\ (\ref{eq:Aprime-4}) to Eq.\
(\ref{eq:Aprime-abs-5}) we have considered the distances in each
arm to be fixed, so that $\overline{x} \rightarrow x$ and
$\overline{x}^{\prime} \rightarrow x^{\prime}$. Then we let
$\tau^{\prime} =\overline{t}+\overline{t}^{\prime}$, and then
$\xi=\tau^{\prime}-\overline{\tau}^{\prime}$ where
$\overline{\tau}^{\prime} = (\alpha_1 x +\alpha_2 x^{\prime})$.
Clearly Eq.\ (\ref{eq:Aprime-abs-5}) is a Gaussian distribution in
$\tau^{\prime}$ with mean $\overline{\tau}^{\prime}$ and variance
\begin{equation}\label{eq:Aprime-variance}
  \sigma_{\tau^{\prime}}^2 = \frac{a_{\phi}^4 + N^2 B^2}{N^2 a_{\phi}^2}
\end{equation}
\begin{equation}\label{eq:Aprime-variance2}
  \sigma_{\tau^{\prime}}^2 = \frac{1+ 4 \sigma_{\phi}^4 N^2 (\beta_1 x +
  \beta_2 x^{\prime})^2}{2 \sigma_{\phi}^2 N^2}.
\end{equation}
Equation (\ref{eq:Aprime-variance2}) is identical in form to Eqs.\
(\ref{eq:A-6-sigma}) and (\ref{eq:A-7-sigma}) which shows that
dispersion cancellation can occur equally well for entangled
states with either correlated or anti-correlated frequencies when
$\beta_1 x= -\beta_2 x^{\prime}$. This is due to the fact that the
dispersive effects are proportional to $\epsilon^2$, which is the
same for correlated or anti-correlated frequencies. However, the
group velocity terms depend on $\epsilon$ itself, with the result
that $\tau^{\prime}$ involves the sum of the detection times
rather than the difference. Thus the detection times are highly
anti-correlated when the frequencies are correlated, whereas the
detection times are correlated in the more usual case where the
frequencies are anti-correlated. This result has different
implications for clock synchronization than before because it is
the sum of the mean arrival times at the two detectors which has a
narrow spread.

In the case where dispersive effects can be neither cancelled nor
neglected, Eq.\ (\ref{eq:Aprime-variance2}) has a limiting value
for large $N$ given by $\sigma_{\tau^{\prime}} \rightarrow
\sqrt{2} \sigma_{\phi} |\beta_1 x + \beta_2 x^{\prime}|$ as
before.

%------------------------------------------------------
\section{Entangled Coherent States \label{sec:coh}}

We have considered so far only Fock states with definite photon
number. In this section it is shown that similar results can be
achieved for frequency-entangled coherent states. The generation
and propagation of entangled coherent states has attracted some
recent interest
\cite{Howell:2000,Chizhov:2001,Filip:2001,Wang:2001}.

A coherent state of frequency $\omega$ is defined as:
\begin{equation}\label{eq:def-coh-state}
  \ket{\mathit{v},\omega} = \exp(-|\mathit{v}|^{2}/2) \sum_{n=0}^{\infty}
  \frac{\mathit{v}^n}{\sqrt{n!}} \ket{n(\omega)}.
\end{equation}
where $\mathit{v}$ is an arbitrary complex parameter
\cite{Glauber:1963}.  This can be expanded using creation
operators as
\begin{equation}\label{eq:def-coh-state-adagger}
  \ket{\mathit{v},\omega} = \exp(-|\mathit{v}|^{2}/2) \sum_{n=0}^{\infty}
  \frac{\mathit{v}^n}{n!} \left(\hat{a}^{\dagger}(\omega)\right)^n \ket{0}.
\end{equation}
A coherent state has the property \cite{note1} that
\begin{equation}\label{eq:property-coh}
 \hat{a}(\omega_i) \ket{\mathit{v},\omega_j} = \mathit{v}
 \, \ket{\mathit{v},\omega_j} \, \delta_{ij}
\end{equation}
which we will make use of below. It follows from Eq.\
(\ref{eq:property-coh}) that the mean number of photons in the
state is $\bra{\mathit{v}} \hat{n} \ket{\mathit{v}} =
\bra{\mathit{v}} \hat{a}^{\dagger} \hat{a} \ket{\mathit{v}} =
|\mathit{v}|^2$.

The original entangled state of Eq.\ (\ref{eq:defPsi}) can be
generalized to
\begin{equation}\label{eq:def-psi-coh}
  \ket{\Psi_{\mathrm{coh}}} = \int d\epsilon \, \phi(\omega_o + \epsilon) \,
  \ket{\mathit{v}, (\omega_o + \epsilon)}_1 \,
  \ket{\mathit{u}, (\omega_o - \epsilon)}_2
\end{equation}
where $|\mathit{v}|$ need not equal $|\mathit{u}|$.
As before, $\braket{A}{A}$ is proportional to the probability of
detecting $N$ photons at times $\{ t_1 \cdots t_N \}$ in detector
1 and $N$ photons at times $\{ t_1^{\prime} \cdots t_N^{\prime}
\}$ in detector 2, where now
\begin{equation}\label{eq:A-coh-1}
\begin{split}
  \ket{A(t_1,\cdots,t_N \, &; \, t_{1}^{\prime},\cdots,t_{N}^{\prime} )} =
  \hat{E}^{(+)}_1 (x ,t_1) \cdots \hat{E}^{(+)}_1 (x ,t_N) \\
  &\times \hat{E}^{(+)}_2 (x^{\prime},t_{1}^{\prime}) \cdots \hat{E}^{(+)}_2 (x^{\prime},t_{N}^{\prime})
  \ket{\Psi_{\mathrm{coh}}} \\
\end{split}
\end{equation}
Inserting the expansion of the electric field operator  from Eq.\
(\ref{eq:E-FTa}) and making use of Eq.\ (\ref{eq:property-coh}) we
have:
\begin{widetext}
\begin{equation}\label{eq:A-coh-2}
  \begin{split}
  \ket{A} = \!\int \!\! d\epsilon \, \phi(\omega_o \!+ \epsilon) & \,
       \mathit{v} \exp[i (k_1 (\omega_o \! + \epsilon) x \! - (\omega_o +
     \epsilon) t_1)]
                 \cdots
       \mathit{v} \exp[i (k_1 (\omega_o \! + \epsilon) x \! - (\omega_o +
     \epsilon) t_N)]  \\
 \times &
       \mathit{u} \exp[i (k_2 (\omega_o \! - \epsilon) x^{\prime} \! - (\omega_o \!-
     \epsilon) t_1^{\prime})]
                 \cdots
       \mathit{u} \exp[i (k_2 (\omega_o \! - \epsilon) x^{\prime} \! - (\omega_o -
     \epsilon) t_N^{\prime})] \\
 \times &
  \ket{\mathit{v}, (\omega_o \! + \epsilon)}_1 \,
  \ket{\mathit{u}, (\omega_o \! - \epsilon)}_2 \\
  \end{split}
\end{equation}
%\end{widetext}
%
%
In computing $\braket{A}{A}$ there arise inner products of the
form $\braket{\mathit{v}^{\prime}, \omega_i}{\mathit{v},
\omega_j}$, but it can be shown \cite{MandelWolf:1995} that
\begin{equation}\label{eq:coh-inner-product}
  \braket{\mathit{v}^{\prime},\omega_i}{\mathit{v},\omega_j} =
  \exp[-|\mathit{v}^{\prime}-\mathit{v}|^{2}/2]
  \exp[(\mathit{v}^{\ast}\mathit{v}^{\prime} -
      \mathit{v} \mathit{v}^{\prime\ast})/2] \delta_{ij}.
\end{equation}
Again, defining $\overline{t}=(1/N) \sum_{j=1}^{N} t_j$ and
similarly for $\overline{t}^{\prime}$, we have:
%\begin{widetext}
\begin{equation}\label{eq:Asq-coh-3}
  |A|^2 = |\mathit{v}|^{2N} |\mathit{u}|^{2N}
  \left| \int \! d\epsilon \, \phi(\omega_o + \epsilon) \,
  \exp[i \epsilon N  (\alpha_1 x - \alpha_2 x^{\prime})
  + i \epsilon^2 N (\beta_1 x +  \beta_2 x^{\prime})
  - i \epsilon N (\overline{t} - \overline{t}^{\prime})]  \right|^{2}
\end{equation}
\end{widetext}
where we have used the dispersive properties of the media.

Equation (\ref{eq:Asq-coh-3}) has the same form as
Eq.(\ref{eq:A-2}) and gives the same results as does the original
entangled Fock state of Eq.\ (\ref{eq:defPsi}), aside from the
overall magnitude of the detection probabilities. The distribution
of arrival times is given by Eqs.\ (\ref{eq:A-5}) through
(\ref{eq:A-7-sigma}). As a result, dispersion cancellation and
non-classical noise reduction can occur just as well for the
entangled coherent states of Eq.\ (\ref{eq:def-psi-coh}) as for
entangled Fock states, which seems somewhat surprising. Equally
surprising is the fact that the dispersion cancellation and noise
reduction are independent of the relative magnitudes of
$|\mathit{u}|$ and $|\mathit{v}|$, provided that equal numbers of
photons are detected in both detectors, as was assumed above.
These results are a direct consequence of the fact that coherent
states are eigenstates of the annihilation operation, as indicated
in Eq.\ (\ref{eq:property-coh}).

%------------------------------------------------
\section{Summary \label{sec:sum}}

     We have generalized nonlocal cancellation of dispersion to
entangled states containing large numbers of phot\-ons.  The same
entangled states were also shown to exhibit a factor of
$1/\sqrt{N}$ reduction in noise below the classical shot noise
limit for timing applications such as clock synchronization
\cite{Jozsa:2000,Giovannetti:2001,Chuang:2000,Burt:2001,Jozsa:2001,%
Yurtsever:2000,Preskill:2000}. Similar results were obtained for
several different types of entangled states, including
anti-correlated states, correlated states, and entangled coherent
states. The fact that effects of this kind can occur for entangled
coherent states shows that these non-classical correlations are a
fundamental result of entanglement and are not limited to number
states.

     Our results also show that relatively small amounts of
dispersion can essentially eliminate the factor of $1/\sqrt{N}$
reduction in noise for timing applications that was proposed by
Giovannetti, Lloyd, and Maccone \cite{Giovannetti:2001}.  This
could have a major impact on practical applications of these
techniques for clock synchronization whenever the photons must
pass through a dispersive medium, such as the Earth's atmosphere.
Dispersion cancellation can, in principle, be used to restore the
$1/\sqrt{N}$ noise reduction, but the effects described here
require that the coefficient of dispersion have the opposite sign
in the two media. This may be the case in some potential
applications, such as clock synchronization using optical fibers,
where the frequencies of the two photons could be chosen to be on
opposite sides of the point of minimum dispersion in the fiber
\cite{Brendel:1998}. There is no obvious way to satisfy this
condition in air, however, which may limit the use of these
techniques to satellite-to-satellite links that do not pass
through the Earth's atmosphere. Other forms of dispersion
cancellation
\cite{SteinbergL:1992,SteinbergA:1992,Giovannetti:2001b} are not
subject to this requirement and may be more useful for free-space
clock synchronization, although they are more restricted with
regard to the optical paths of the photons and in other respects.

     Regardless of any potential practical applications, these results
show that there is a close connection between non-classical noise
reduction and nonlocal cancellation of dispersion, which we have
generalized to states containing large numbers of photons.

\begin{acknowledgments}
The authors would like to acknowledge valuable discussions with
Jonathan P. Dowling, Kevin J. McCann, and Todd B. Pittman. This
work was supported by the National Reconnaissance Office and the
Office of Naval Research.
\end{acknowledgments}

\bibliography{nphot-3}

\begin{thebibliography}{38}
\expandafter\ifx\csname natexlab\endcsname\relax\def\natexlab#1{#1}\fi
\expandafter\ifx\csname bibnamefont\endcsname\relax
  \def\bibnamefont#1{#1}\fi
\expandafter\ifx\csname bibfnamefont\endcsname\relax
  \def\bibfnamefont#1{#1}\fi
\expandafter\ifx\csname citenamefont\endcsname\relax
  \def\citenamefont#1{#1}\fi
\expandafter\ifx\csname url\endcsname\relax
  \def\url#1{\texttt{#1}}\fi
\expandafter\ifx\csname urlprefix\endcsname\relax\def\urlprefix{URL }\fi
\providecommand{\bibinfo}[2]{#2}
\providecommand{\eprint}[2][]{\url{#2}}

\bibitem[{\citenamefont{Franson}(1992)}]{Franson:1992}
\bibinfo{author}{\bibfnamefont{J.~D.} \bibnamefont{Franson}},
  \bibinfo{journal}{Phys.\ Rev.\ A} \textbf{\bibinfo{volume}{45}},
  \bibinfo{pages}{3126} (\bibinfo{year}{1992}).

\bibitem[{\citenamefont{Steinberg
  et~al.}(1992{\natexlab{a}})\citenamefont{Steinberg, Kwiat, and
  Chiao}}]{SteinbergL:1992}
\bibinfo{author}{\bibfnamefont{A.~M.} \bibnamefont{Steinberg}},
  \bibinfo{author}{\bibfnamefont{P.~G.} \bibnamefont{Kwiat}}, \bibnamefont{and}
  \bibinfo{author}{\bibfnamefont{R.~Y.} \bibnamefont{Chiao}},
  \bibinfo{journal}{Phys.\ Rev.\ Lett.} \textbf{\bibinfo{volume}{68}},
  \bibinfo{pages}{2421} (\bibinfo{year}{1992}{\natexlab{a}}).

\bibitem[{\citenamefont{Steinberg
  et~al.}(1992{\natexlab{b}})\citenamefont{Steinberg, Kwiat, and
  Chiao}}]{SteinbergA:1992}
\bibinfo{author}{\bibfnamefont{A.~M.} \bibnamefont{Steinberg}},
  \bibinfo{author}{\bibfnamefont{P.~G.} \bibnamefont{Kwiat}}, \bibnamefont{and}
  \bibinfo{author}{\bibfnamefont{R.~Y.} \bibnamefont{Chiao}},
  \bibinfo{journal}{Phys.\ Rev.\ A} \textbf{\bibinfo{volume}{45}},
  \bibinfo{pages}{6659} (\bibinfo{year}{1992}{\natexlab{b}}).

\bibitem[{\citenamefont{Giovannetti
  et~al.}(2001{\natexlab{a}})\citenamefont{Giovannetti, Lloyd, Maccone, and
  Wong}}]{Giovannetti:2001b}
\bibinfo{author}{\bibfnamefont{V.}~\bibnamefont{Giovannetti}},
  \bibinfo{author}{\bibfnamefont{S.}~\bibnamefont{Lloyd}},
  \bibinfo{author}{\bibfnamefont{L.}~\bibnamefont{Maccone}}, \bibnamefont{and}
  \bibinfo{author}{\bibfnamefont{F.~N.~C.} \bibnamefont{Wong}},
  \bibinfo{journal}{Phys.\ Rev.\ Lett.} \textbf{\bibinfo{volume}{87}},
  \bibinfo{pages}{117902} (\bibinfo{year}{2001}{\natexlab{a}}).

\bibitem[{\citenamefont{Larchuk et~al.}(1995)\citenamefont{Larchuk, Teich, and
  Saleh}}]{Larchuk:1995}
\bibinfo{author}{\bibfnamefont{T.~S.} \bibnamefont{Larchuk}},
  \bibinfo{author}{\bibfnamefont{M.~C.} \bibnamefont{Teich}}, \bibnamefont{and}
  \bibinfo{author}{\bibfnamefont{B.~E.~A.} \bibnamefont{Saleh}},
  \bibinfo{journal}{Phys.\ Rev.\ A} \textbf{\bibinfo{volume}{52}},
  \bibinfo{pages}{4145} (\bibinfo{year}{1995}).

\bibitem[{\citenamefont{Agarwal and Gupta}(1994)}]{Agarwal:1994}
\bibinfo{author}{\bibfnamefont{G.~S.} \bibnamefont{Agarwal}} \bibnamefont{and}
  \bibinfo{author}{\bibfnamefont{S.~D.} \bibnamefont{Gupta}},
  \bibinfo{journal}{Phys.\ Rev.\ A} \textbf{\bibinfo{volume}{49}},
  \bibinfo{pages}{3954} (\bibinfo{year}{1994}).

\bibitem[{\citenamefont{Jeffers and Barnett}(1993)}]{Jeffers:1993}
\bibinfo{author}{\bibfnamefont{J.}~\bibnamefont{Jeffers}} \bibnamefont{and}
  \bibinfo{author}{\bibfnamefont{S.~M.} \bibnamefont{Barnett}},
  \bibinfo{journal}{Phys.\ Rev.\ A} \textbf{\bibinfo{volume}{47}},
  \bibinfo{pages}{3291} (\bibinfo{year}{1993}).

\bibitem[{\citenamefont{Jozsa et~al.}(2000)\citenamefont{Jozsa, Abrams,
  Dowling, and Williams}}]{Jozsa:2000}
\bibinfo{author}{\bibfnamefont{R.}~\bibnamefont{Jozsa}},
  \bibinfo{author}{\bibfnamefont{D.~S.} \bibnamefont{Abrams}},
  \bibinfo{author}{\bibfnamefont{J.~P.} \bibnamefont{Dowling}},
  \bibnamefont{and} \bibinfo{author}{\bibfnamefont{C.~P.}
  \bibnamefont{Williams}}, \bibinfo{journal}{Phys.\ Rev.\ Lett.}
  \textbf{\bibinfo{volume}{85}}, \bibinfo{pages}{2010} (\bibinfo{year}{2000}).

\bibitem[{\citenamefont{Giovannetti
  et~al.}(2001{\natexlab{b}})\citenamefont{Giovannetti, Lloyd, and
  Maccone}}]{Giovannetti:2001}
\bibinfo{author}{\bibfnamefont{V.}~\bibnamefont{Giovannetti}},
  \bibinfo{author}{\bibfnamefont{S.}~\bibnamefont{Lloyd}}, \bibnamefont{and}
  \bibinfo{author}{\bibfnamefont{L.}~\bibnamefont{Maccone}},
  \bibinfo{journal}{Nature} \textbf{\bibinfo{volume}{412}},
  \bibinfo{pages}{417} (\bibinfo{year}{2001}{\natexlab{b}}).

\bibitem[{\citenamefont{Chuang}(2000)}]{Chuang:2000}
\bibinfo{author}{\bibfnamefont{I.~L.} \bibnamefont{Chuang}},
  \bibinfo{journal}{Phys.\ Rev.\ Lett.} \textbf{\bibinfo{volume}{85}},
  \bibinfo{pages}{2006} (\bibinfo{year}{2000}).

\bibitem[{\citenamefont{Burt et~al.}(2001)\citenamefont{Burt, Ekstrom, and
  Swanson}}]{Burt:2001}
\bibinfo{author}{\bibfnamefont{E.~A.} \bibnamefont{Burt}},
  \bibinfo{author}{\bibfnamefont{C.~R.} \bibnamefont{Ekstrom}},
  \bibnamefont{and} \bibinfo{author}{\bibfnamefont{T.~B.}
  \bibnamefont{Swanson}}, \bibinfo{journal}{Phys.\ Rev.\ Lett.}
  \textbf{\bibinfo{volume}{87}}, \bibinfo{pages}{129801}
  (\bibinfo{year}{2001}), \bibinfo{note}{quant-ph/0007030}.

\bibitem[{\citenamefont{Jozsa et~al.}(2001)\citenamefont{Jozsa, Abrams,
  Dowling, and Williams}}]{Jozsa:2001}
\bibinfo{author}{\bibfnamefont{R.}~\bibnamefont{Jozsa}},
  \bibinfo{author}{\bibfnamefont{D.~S.} \bibnamefont{Abrams}},
  \bibinfo{author}{\bibfnamefont{J.~P.} \bibnamefont{Dowling}},
  \bibnamefont{and} \bibinfo{author}{\bibfnamefont{C.~P.}
  \bibnamefont{Williams}}, \bibinfo{journal}{Phys.\ Rev.\ Lett.}
  \textbf{\bibinfo{volume}{87}}, \bibinfo{pages}{129802}
  (\bibinfo{year}{2001}).

\bibitem[{\citenamefont{Yurtsever and Dowling}(2000)}]{Yurtsever:2000}
\bibinfo{author}{\bibfnamefont{U.}~\bibnamefont{Yurtsever}} \bibnamefont{and}
  \bibinfo{author}{\bibfnamefont{J.~P.} \bibnamefont{Dowling}},
  \emph{\bibinfo{title}{A {L}orentz-invariant look at quantum clock
  synchronization protocols based on distributed engtanglement}}
  (\bibinfo{year}{2000}), \eprint{quant-ph/0010097}.

\bibitem[{\citenamefont{Preskill}(2000)}]{Preskill:2000}
\bibinfo{author}{\bibfnamefont{J.}~\bibnamefont{Preskill}},
  \emph{\bibinfo{title}{Quantum clock synchronization and quantum error
  correction}} (\bibinfo{year}{2000}), \eprint{quant-ph/0010098}.

\bibitem[{\citenamefont{Giovannetti et~al.}(2002)\citenamefont{Giovannetti,
  Lloyd, and Maccone}}]{Giovannetti:2002}
\bibinfo{author}{\bibfnamefont{V.}~\bibnamefont{Giovannetti}},
  \bibinfo{author}{\bibfnamefont{S.}~\bibnamefont{Lloyd}}, \bibnamefont{and}
  \bibinfo{author}{\bibfnamefont{L.}~\bibnamefont{Maccone}},
  \bibinfo{journal}{Phys. Rev. A} \textbf{\bibinfo{volume}{65}},
  \bibinfo{pages}{022309} (\bibinfo{year}{2002}).

\bibitem[{not()}]{note1}
\bibinfo{note}{It would be more precise to use boundary conditions that are
  periodic in a length $L$, which gives discrete frequencies and a commutator
  of the form ${{ [\hat{a}_{\ell} (\omega_{i}) \, , \,
  \hat{a}_{k}^{\dagger}(\omega_{j})] = \delta_{\ell k} \, \delta_{ij} }}$.
  After using the commutation relations, the remaining sums would reduce to the
  integrals shown in the text in the limit of large $L$, as usual.}

\bibitem[{\citenamefont{Weinfurter and {\.{Z}}ukowski}(2001)}]{Weinfurter:2001}
\bibinfo{author}{\bibfnamefont{H.}~\bibnamefont{Weinfurter}} \bibnamefont{and}
  \bibinfo{author}{\bibfnamefont{M.}~\bibnamefont{{\.{Z}}ukowski}},
  \bibinfo{journal}{Phys.\ Rev.\ A} \textbf{\bibinfo{volume}{64}},
  \bibinfo{pages}{010102(R)} (\bibinfo{year}{2001}).

\bibitem[{\citenamefont{Pan et~al.}(2001)\citenamefont{Pan, Daniell, Gasparoni,
  Weihs, and Zeilinger}}]{Pan:2001}
\bibinfo{author}{\bibfnamefont{J.-W.} \bibnamefont{Pan}},
  \bibinfo{author}{\bibfnamefont{M.}~\bibnamefont{Daniell}},
  \bibinfo{author}{\bibfnamefont{S.}~\bibnamefont{Gasparoni}},
  \bibinfo{author}{\bibfnamefont{G.}~\bibnamefont{Weihs}}, \bibnamefont{and}
  \bibinfo{author}{\bibfnamefont{A.}~\bibnamefont{Zeilinger}},
  \bibinfo{journal}{Phys.\ Rev.\ Lett.} \textbf{\bibinfo{volume}{86}},
  \bibinfo{pages}{4435} (\bibinfo{year}{2001}).

\bibitem[{\citenamefont{Ou et~al.}(1999)\citenamefont{Ou, Rhee, and
  Wang}}]{Ou:1999}
\bibinfo{author}{\bibfnamefont{Z.~Y.} \bibnamefont{Ou}},
  \bibinfo{author}{\bibfnamefont{J.-K.} \bibnamefont{Rhee}}, \bibnamefont{and}
  \bibinfo{author}{\bibfnamefont{L.~J.} \bibnamefont{Wang}},
  \bibinfo{journal}{Phys.\ Rev.\ Lett.} \textbf{\bibinfo{volume}{83}},
  \bibinfo{pages}{959} (\bibinfo{year}{1999}).

\bibitem[{\citenamefont{Lee et~al.}(2001)\citenamefont{Lee, Kok, Cerf, and
  Dowling}}]{Lee:2001}
\bibinfo{author}{\bibfnamefont{H.}~\bibnamefont{Lee}},
  \bibinfo{author}{\bibfnamefont{P.}~\bibnamefont{Kok}},
  \bibinfo{author}{\bibfnamefont{N.~J.} \bibnamefont{Cerf}}, \bibnamefont{and}
  \bibinfo{author}{\bibfnamefont{J.~P.} \bibnamefont{Dowling}},
  \emph{\bibinfo{title}{Linear optics and projective measurements alone suffice
  to create large-photon-number path entanglement}} (\bibinfo{year}{2001}),
  \eprint{quant-ph/0109080}.

\bibitem[{\citenamefont{Kok et~al.}(2001)\citenamefont{Kok, Lee, and
  Dowling}}]{Kok:2001}
\bibinfo{author}{\bibfnamefont{P.}~\bibnamefont{Kok}},
  \bibinfo{author}{\bibfnamefont{H.}~\bibnamefont{Lee}}, \bibnamefont{and}
  \bibinfo{author}{\bibfnamefont{J.~P.} \bibnamefont{Dowling}},
  \emph{\bibinfo{title}{The creation of large photon-number path entanglement
  conditioned on photodetection}} (\bibinfo{year}{2001}),
  \eprint{quant-ph/0112002}.

\bibitem[{\citenamefont{Fiur{\'{a}}{\v{s}}ek}(2001)}]{Fiurasek:2001}
\bibinfo{author}{\bibfnamefont{J.}~\bibnamefont{Fiur{\'{a}}{\v{s}}ek}},
  \emph{\bibinfo{title}{Conditional generation of n-photon entangled states of
  light}} (\bibinfo{year}{2001}), \eprint{quant-ph/0110138}.

\bibitem[{\citenamefont{Lamas-Linares et~al.}(2001)\citenamefont{Lamas-Linares,
  Howell, and Bouwmeester}}]{LLinares:2001}
\bibinfo{author}{\bibfnamefont{A.}~\bibnamefont{Lamas-Linares}},
  \bibinfo{author}{\bibfnamefont{J.~C.} \bibnamefont{Howell}},
  \bibnamefont{and}
  \bibinfo{author}{\bibfnamefont{D.}~\bibnamefont{Bouwmeester}},
  \bibinfo{journal}{Nature} \textbf{\bibinfo{volume}{412}},
  \bibinfo{pages}{887} (\bibinfo{year}{2001}).

\bibitem[{\citenamefont{Boto et~al.}(2000)\citenamefont{Boto, Kok, Abrams,
  Braunstein, Williams, and Dowling}}]{Boto:2000}
\bibinfo{author}{\bibfnamefont{A.~N.} \bibnamefont{Boto}},
  \bibinfo{author}{\bibfnamefont{P.}~\bibnamefont{Kok}},
  \bibinfo{author}{\bibfnamefont{D.~S.} \bibnamefont{Abrams}},
  \bibinfo{author}{\bibfnamefont{S.~L.} \bibnamefont{Braunstein}},
  \bibinfo{author}{\bibfnamefont{C.~P.} \bibnamefont{Williams}},
  \bibnamefont{and} \bibinfo{author}{\bibfnamefont{J.~P.}
  \bibnamefont{Dowling}}, \bibinfo{journal}{Phys.\ Rev.\ Lett.}
  \textbf{\bibinfo{volume}{85}}, \bibinfo{pages}{2733} (\bibinfo{year}{2000}).

\bibitem[{\citenamefont{Brendel et~al.}(1998)\citenamefont{Brendel, Zbinden,
  and Gisin}}]{Brendel:1998}
\bibinfo{author}{\bibfnamefont{J.}~\bibnamefont{Brendel}},
  \bibinfo{author}{\bibfnamefont{H.}~\bibnamefont{Zbinden}}, \bibnamefont{and}
  \bibinfo{author}{\bibfnamefont{N.}~\bibnamefont{Gisin}},
  \bibinfo{journal}{Opt.\ Commun.} \textbf{\bibinfo{volume}{151}},
  \bibinfo{pages}{35} (\bibinfo{year}{1998}).

\bibitem[{\citenamefont{Walmsley et~al.}(2001)\citenamefont{Walmsley, Waxer,
  and Dorrer}}]{Walmsley:2001}
\bibinfo{author}{\bibfnamefont{I.}~\bibnamefont{Walmsley}},
  \bibinfo{author}{\bibfnamefont{L.}~\bibnamefont{Waxer}}, \bibnamefont{and}
  \bibinfo{author}{\bibfnamefont{C.}~\bibnamefont{Dorrer}},
  \bibinfo{journal}{Rev.\ Sci.\ Instr.} \textbf{\bibinfo{volume}{72}},
  \bibinfo{pages}{1} (\bibinfo{year}{2001}).

\bibitem[{\citenamefont{Edl{\'{e}}n}(1966)}]{Edlen:1966}
\bibinfo{author}{\bibfnamefont{B.}~\bibnamefont{Edl{\'{e}}n}},
  \bibinfo{journal}{Metrologia} \textbf{\bibinfo{volume}{2}},
  \bibinfo{pages}{71} (\bibinfo{year}{1966}).

\bibitem[{\citenamefont{Owens}(1967)}]{Owens:1967}
\bibinfo{author}{\bibfnamefont{J.~C.} \bibnamefont{Owens}},
  \bibinfo{journal}{Appl.\ Opt.} \textbf{\bibinfo{volume}{6}},
  \bibinfo{pages}{51} (\bibinfo{year}{1967}).

\bibitem[{\citenamefont{Branning et~al.}(1999)\citenamefont{Branning, Grice,
  Erdmann, and Walmsley}}]{Branning:1999}
\bibinfo{author}{\bibfnamefont{D.}~\bibnamefont{Branning}},
  \bibinfo{author}{\bibfnamefont{W.~P.} \bibnamefont{Grice}},
  \bibinfo{author}{\bibfnamefont{R.}~\bibnamefont{Erdmann}}, \bibnamefont{and}
  \bibinfo{author}{\bibfnamefont{I.~A.} \bibnamefont{Walmsley}},
  \bibinfo{journal}{Phys. Rev. Lett.} \textbf{\bibinfo{volume}{83}},
  \bibinfo{pages}{955} (\bibinfo{year}{1999}).

\bibitem[{\citenamefont{Erdmann et~al.}(2000)\citenamefont{Erdmann, Branning,
  Grice, and Walmsley}}]{Erdmann:2000}
\bibinfo{author}{\bibfnamefont{R.}~\bibnamefont{Erdmann}},
  \bibinfo{author}{\bibfnamefont{D.}~\bibnamefont{Branning}},
  \bibinfo{author}{\bibfnamefont{W.}~\bibnamefont{Grice}}, \bibnamefont{and}
  \bibinfo{author}{\bibfnamefont{I.~A.} \bibnamefont{Walmsley}},
  \bibinfo{journal}{Phys. Rev. A} \textbf{\bibinfo{volume}{62}},
  \bibinfo{pages}{053810} (\bibinfo{year}{2000}).

\bibitem[{\citenamefont{Banaszek et~al.}(2001)\citenamefont{Banaszek, U'Ren,
  and Walmsley}}]{Banaszek:2001}
\bibinfo{author}{\bibfnamefont{K.}~\bibnamefont{Banaszek}},
  \bibinfo{author}{\bibfnamefont{A.~B.} \bibnamefont{U'Ren}}, \bibnamefont{and}
  \bibinfo{author}{\bibfnamefont{I.~A.} \bibnamefont{Walmsley}},
  \bibinfo{journal}{Opt. Lett.} \textbf{\bibinfo{volume}{26}},
  \bibinfo{pages}{1367} (\bibinfo{year}{2001}).

\bibitem[{\citenamefont{Campos et~al.}(1990)\citenamefont{Campos, Saleh, and
  Teich}}]{Campos:1990}
\bibinfo{author}{\bibfnamefont{R.~A.} \bibnamefont{Campos}},
  \bibinfo{author}{\bibfnamefont{B.~E.~A.} \bibnamefont{Saleh}},
  \bibnamefont{and} \bibinfo{author}{\bibfnamefont{M.~C.} \bibnamefont{Teich}},
  \bibinfo{journal}{Phys.\ Rev.\ A} \textbf{\bibinfo{volume}{42}},
  \bibinfo{pages}{4127} (\bibinfo{year}{1990}).

\bibitem[{\citenamefont{Howell and Yeazell}(2000)}]{Howell:2000}
\bibinfo{author}{\bibfnamefont{J.~C.} \bibnamefont{Howell}} \bibnamefont{and}
  \bibinfo{author}{\bibfnamefont{J.~A.} \bibnamefont{Yeazell}},
  \bibinfo{journal}{Phys.\ Rev.\ A} \textbf{\bibinfo{volume}{62}},
  \bibinfo{pages}{012102} (\bibinfo{year}{2000}).

\bibitem[{\citenamefont{Chizhov et~al.}(2001)\citenamefont{Chizhov, Schmidt,
  Kn{\"{o}}ll, and Welsch}}]{Chizhov:2001}
\bibinfo{author}{\bibfnamefont{A.~V.} \bibnamefont{Chizhov}},
  \bibinfo{author}{\bibfnamefont{E.}~\bibnamefont{Schmidt}},
  \bibinfo{author}{\bibfnamefont{L.}~\bibnamefont{Kn{\"{o}}ll}},
  \bibnamefont{and} \bibinfo{author}{\bibfnamefont{D.-G.}
  \bibnamefont{Welsch}}, \bibinfo{journal}{J.\ Opt.\ B: Quantum Semiclass.\
  Opt.} \textbf{\bibinfo{volume}{3}}, \bibinfo{pages}{77}
  (\bibinfo{year}{2001}).

\bibitem[{\citenamefont{Filip et~al.}(2001)\citenamefont{Filip,
  {\v{R}}eh{\'{a}}{\v{c}}ek, and Du{\v{s}}ek}}]{Filip:2001}
\bibinfo{author}{\bibfnamefont{R.}~\bibnamefont{Filip}},
  \bibinfo{author}{\bibfnamefont{J.}~\bibnamefont{{\v{R}}eh{\'{a}}{\v{c}}ek}},
  \bibnamefont{and}
  \bibinfo{author}{\bibfnamefont{M.}~\bibnamefont{Du{\v{s}}ek}},
  \bibinfo{journal}{J.\ Opt.\ B: Quantum Semiclass.\ Opt.}
  \textbf{\bibinfo{volume}{3}}, \bibinfo{pages}{341} (\bibinfo{year}{2001}).

\bibitem[{\citenamefont{Wang et~al.}(2001)\citenamefont{Wang, Hong, and
  Friberg}}]{Wang:2001}
\bibinfo{author}{\bibfnamefont{L.~J.} \bibnamefont{Wang}},
  \bibinfo{author}{\bibfnamefont{C.~K.} \bibnamefont{Hong}}, \bibnamefont{and}
  \bibinfo{author}{\bibfnamefont{S.~R.} \bibnamefont{Friberg}},
  \bibinfo{journal}{J.\ Opt.\ B: Quantum Semiclass.\ Opt.}
  \textbf{\bibinfo{volume}{3}}, \bibinfo{pages}{346} (\bibinfo{year}{2001}).

\bibitem[{\citenamefont{Glauber}(1963)}]{Glauber:1963}
\bibinfo{author}{\bibfnamefont{R.~J.} \bibnamefont{Glauber}},
  \bibinfo{journal}{Phys.\ Rev.} \textbf{\bibinfo{volume}{131}},
  \bibinfo{pages}{2766} (\bibinfo{year}{1963}).

\bibitem[{\citenamefont{Mandel and Wolf}(1995)}]{MandelWolf:1995}
\bibinfo{author}{\bibfnamefont{L.}~\bibnamefont{Mandel}} \bibnamefont{and}
  \bibinfo{author}{\bibfnamefont{E.}~\bibnamefont{Wolf}},
  \emph{\bibinfo{title}{Optical coherence and quantum optics}}
  (\bibinfo{publisher}{Cambridge University Press},
  \bibinfo{address}{Cambridge, England}, \bibinfo{year}{1995}).

\end{thebibliography}

\end{document}